# Comparative Study of MAC Layer Protocols in Wireless Sensor Networks: A Survey

Rahul R Lanjewar[#1], Dr D S Adane[*2]

[#]*Student, Department of Computer Science and Engineering*
*Ramdeobaba College of Engineering and Management Nagpur, India*
[*]*HoD, Department of Information Technology*
*Ramdeobaba College of Engineering and Management Nagpur, India*

*Abstract*— Wireless Sensor Networks (WSN) is the collection of many small size low cost, battery operated sensor nodes distributed over the targeted region to collect the information of interest. We can say these networks can be a fruitful solution for many applications such as target tracking, intrusion detection etc. Whenever we are talking about MAC layer protocols we need to give stress on energy efficiency factor. There are few other issues like high throughput and delay. In the early development stages, designers were mostly concerned with energy efficiency because sensor nodes are usually limited in power supply. Recently, new protocols are being developed to provide multi-task support and efficient delivery of bursty traffic. Therefore, research attention has turned back to throughput and delay. Designing an efficient MAC layer protocol is an important task as it coordinates all the nodes to the share the wireless medium. In [3] classification of MAC layer protocols is carried out in four categories viz. Asynchronous, Synchronous, Frame-Slotted and Multichannel. We are carrying the same classification. In our survey we have compared different MAC protocols in terms of energy efficiency, data delivery mechanisms and overhead to maintain a protocol along with their advantages and disadvantages.

## I. INTRODUCTION

Wireless Sensor Networks consist of large number of small size sensor nodes having low cost distributed over a targeted region to collect the information by doing wireless communication. These small sensing nodes consist of battery for energy and transceiver for receiving and sending signals or data from one node CPU for data processing, memory for data storage [1]. These networks are used in many applications like target tracking, environmental monitoring, disaster relief, field survey, intrusion detection etc. But low battery resources have engaged researchers to innovate new techniques to achieve more efficiency. Energy efficiency is a fundamental criterion in the design of WSN MAC protocols because Sensor nodes in a network has to work independently for months or even for years. A major power consuming component of a sensor node is the radio, which is controlled by the MAC protocol. Hence by designing efficient MAC protocols we can increase the lifetime of wireless sensor networks. In addition, the MAC layer controls how nodes share the wireless medium. An efficient MAC protocol can reduce collisions and increase the achievable throughput, providing flexibility for various applications [1][10].

MAC layer protocols must be scalable and adaptable means they have to adapt themselves in continuously changing environment such as node density, network size or topology [3][10][12]. We have also considered the factors like latency, bandwidth utilization, fairness, throughput for the comparison etc. These factors have secondary priority in terms of MAC protocols in wireless Sensor Networks. In [3] they have classified MAC protocols in 4 categories such as Asynchronous, Synchronous, Frame-Slotted and Multichannel protocols. We are going to compare them by considering the same classification. Each category has some pros and cons in terms of the MAC protocols used in that category. Multichannel MAC layer protocols are the recent trends in MAC protocols for Wireless Sensor Networks.

The rest of the paper is organized as follows. In Section II we are discussing asynchronous protocols followed by the synchronous protocols in section III. Section IV will cover Frame-Slotted and in section V Multichannel MAC protocols were discussed. In section VI observation and finally we have made conclusion in section VII.

## II. ASYNCHRONOUS PROTOCOLS

Instead of synchronizing all its neighbors the sensor node itself maintain its own schedule to process the information. With this technique it can saves its cost of synchronizing and will be able to achieve low duty cycle. A node cannot be active for long time so it has to wake up periodically to check for data of interest. For that purpose a preamble sampling [3] technique is used in which data sent by the sender along with the long preamble to the intended receivers. But due to long preamble a problem of over utilization of channel occurs which will leads to limited throughput. There are several techniques discussed in [3][4] which shows the adaptations carried out with the size of preamble and how they tried to maximize the throughput.

There are various asynchronous MAC protocols which have mostly an application oriented designs. B-MAC (Berkeley MAC) [2][3][18] uses a preamble sampling to reduce the idle listening problem which a major source of energy wastage. Also to avoid collisions CSMA with preamble sampling performs CCA (clear channel assessment) before transmitting a preamble. B-MAC [2] [3][18] performs an outlier detection to improve the quality of CCA. This special technique of preamble sampling is known as Low Power Listening (LPL). Still the hidden terminal problem has not been solved because preamble transmission of one node may collide with the data transmission of neighbor. So to avoid this STEM (Sparse Topology and Energy Management) [3] protocol is introduced with two radios one for data and another for wake up channel. STEM has two categories STEM-T (STEM Tone) and STEM-B (STEM Beacon). STEM-B is less energy efficient than STEM-T as channel sampling period must be longer than the wake-up tone in STEM-T.

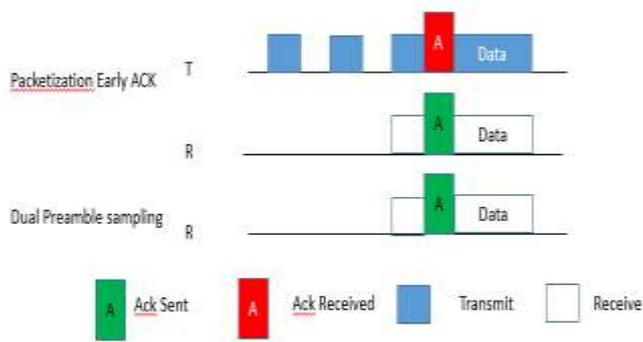

Fig. 1 Continuous Preamble Sampling [3]

These two protocols were facing the problem of collision, hidden terminal problems due to long preamble. So in [3] they have discussed the technique to reduce the length of preamble by packetization with their respective protocols. Protocols like ENBMAC (Enhanced MAC) [3] is used to detect the overhearing problem by including timing information about when the data transmission begins. On the basis of gap in the chunks of packets they categorize this technique into two parts Continuous preamble sampling shown in Fig. 1 and Strobed preamble sampling shown in Fig. 2. According to that time a node will decide to stay active or in sleep mode. X-MAC[3][5] protocol proposed the use of a series of short preamble packets with the destination address embedded in the packet. It is a kind of strobed preamble sampling protocol in which after sending first preamble and successfully received at the receiver ACK will be sent and we can send data immediately. Hence we can reduce the data transmission delay and make energy efficient protocol also we can avoid the idle listening and overhearing by the neighboring nodes. Speck MAC [5] protocol is a kind of continuous preamble sampling type. It includes redundant short packet transmission with embedded destination address. Two types of Speck MAC first is Speck MAC-B (Speck MAC-Back off) and Speck MAC-D (Speck MAC-Data). Speck MAC-D is energy efficient in broadcast transmission while Speck MAC-B is useful in unicast data transmission. So by combining the advantages of both protocols Speck MAC-H [5] was introduced.

Nowadays researchers are working at receiver side to reduce collision. RC-MAC [3] protocol coordinates multiple sender's transmissions by piggybacking a scheduling message to an ACK. Another protocol which works to estimate the wake-up time of receiver is PW-MAC (Predictive wakeup MAC). It introduces a method to predict the target receiver's wake-up time so that a sender only needs to wake up slightly before the target receiver. Also it has advantage over the Wise-MAC [3] that it has pseudo-random schedule which avoids collisions.

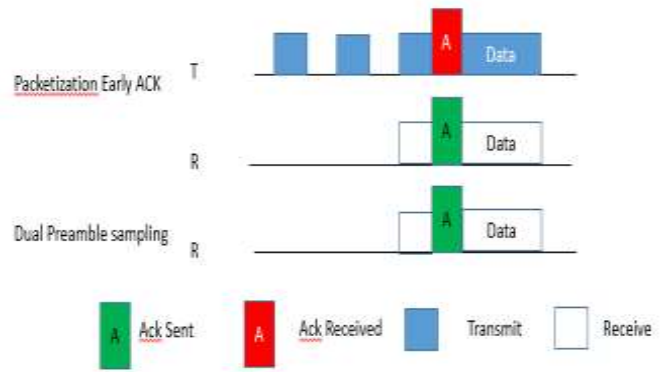

Fig. 2 Strobed Preamble Sampling [3]

III. SYNCHRONOUS PROTOCOLS

Synchronous MAC protocols [3] make clusters from many nodes to set up a common active/sleep schedule within a cluster. But if a particular node belongs to more than one cluster then it has to save multiple schedules which prone to conflicts. In Fig. 3 the node which belongs to both the cluster has to maintain 2 schedules. We have to bare additional cost for synchronization overhead. Here nodes listen for the channel to get a schedule if schedule is not available then it determines its next wake up time and broadcasts its own schedule. If it receives the schedule of neighboring cluster before broadcasting our own. It can follow the schedule of neighbors. Also node has capability to accept one or more schedules and can be a bridge between two clusters. Synchronous MAC protocols require local time synchronization mechanism. These kind of MAC protocols mostly focus on throughput and delay as establishing a connection was not a big issue. They are more accurate for applications of periodic traffic where the wake-up schedule is easy to determine. S-MAC (Sensor MAC) [2] [3] protocol is one of the classical protocol in which clusters have to adopt schedules at the beginning of SYNC period. There are also two periods such as DATA and SLEEP. Nodes can communicate with the exchange of RTS (Request to send) and CTS (Clear to send) during the DATA period. Nodes which are not involved in communication will go in SLEEP period. S-MAC is further extended to several protocols such as T-MAC (Timeout MAC). This protocol used in a future request to send technique. In order to achieve optimal active periods under varying traffic T-MAC [3] sends FRTS (Future request to send) packet to the node and tells the node that channel is not accessible currently. But still it was unable to overcome the problem of overhearing as a node has to stay awake during data transmission. This protocol further extended to RMAC protocol [3]. It is used to reduce latency in multi-hop forwarding. Instead of exchanging data during the DATA period, a control frame called Pioneer Frame (PION) is forwarded by multiple hops. PION frame works as RTS and CTS. DW-MAC protocol has SCH (Scheduling frame) instead of PION and rest of the mechanism is same as that of RMAC.

TABLE I
DIFFERENT ASYNCHRONOUS MAC PROTOCOLS WITH THEIR TECHNIQUES AND ENERGY REQUIREMENT [2][3].

| Protocols | Technique | Advantage | Disadvantages |
|---|---|---|---|
| B-MAC (6 joules) | Low power listening | Frames not required(RTS,CTS) delay tolerant | Long preamble creates large overhead |
| X-MAC (70% of duty cycle) | Reduce preamble length | More energy efficient and lower latency operation. | Mistakenly data transmission by neighbor after seeing gaps in packets |
| Speck MAC | Continuous preamble sampling | Reduces the energy at receiver. | Waste in transmission power due to redundancy |
| RC-MAC | Reduce collision in Receiver initiated transmission | High throughput even in heavy traffic loads. | Delay may increase. |
| PW-MAC (10% of duty cycle) | Estimate wakeup timer at receiver | Due to pseudo-random schedules avoid collisions, low delay. | Overhead created by beacons and idle listening. |
| DPS-MAC | LPL and Short strobed preamble | High Energy Efficient for low traffic application Reduces Idle listening. | Sensitive to switching time of radio which affects the size of short preamble. |

D-MAC (Dynamic MAC) [3] protocol is data gathering oriented protocol in which tree data structure is the basis for this protocol. As tree consists of nodes which are arrange in sequential order from leaf nodes to the root. Here by using this protocol energy is saved by arranging nodes in a particular sequence towards the root or sink node. This leads to low delay. In order to increase the number of active slots, a data prediction mechanism and a More-to-Send (MTS) notification are introduced. Q-MAC protocol also works same only the difference is that active periods are shifted in a way that facilitates downlink traffic. Till now we have discussed some synchronous MAC protocols with adaptability of nodes in a cluster. With the help of SCP-MAC (Scheduled Channel Polling MAC) a small preamble can wake up the receiver. In order to reduce multi-hop latency SCP-MAC develops an adaptive channel polling mechanism to add additional polling slots along the path.

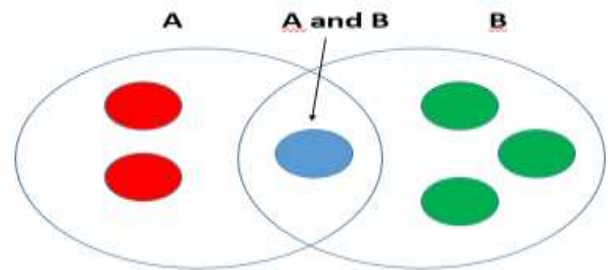

Fig. 3 Nodes in one Cluster have same schedule

TABLE II
DIFFERENT SYNCHRONOUS MAC PROTOCOLS [2][3] [6].

| Protocols | Technique | Advantages | Disadvantages |
|---|---|---|---|
| S-MAC | Adaptive Listening | Low energy consumption, Adaptive to changes in topology. | Need to maintain loose sync. , RTS,CTS increase energy |
| T-MAC | Future request to send | Achieve optimal active period | Still overhearing problem |
| RMAC | Shifting data transmission to sleep period | No overhearing problem | Two hidden terminals may cause collision. |
| D-MAC | Staggered Schedule | Low delay, flexible to increase active slots | Incurs long idle listening, contention may occur at sink |
| SCP-MAC | Adaptive duty cycle | Less Schedule maintenance | Synchronization overhead, listen interval is too long in existing protocol |

IV. FRAME-SLOTTED MAC PROTOCOLS

Frame-Slotted MAC protocols [3] are derived from Time Division Multiple Access (TDMA) protocols. TDMA is also used in synchronous protocols but if active periods of two clusters are overlap then there is a collision. Hence TDMA follows global time synchronization rather than local time synchronization. But still Frame-Slotted MAC protocols will not introduce any additional overhead for a particular application. Frame-slotted MAC protocols are also popular in small scale networks. A special case of WSN is the wireless body area network (WBAN). To provide high throughput, frame-slotted mechanisms allocate time slots in a way that no two nodes within the two-hop communication neighborhood are assigned to the same slot. With this the problem of collision can overcome but hidden terminal problem is still remains a concern. When very few nodes have to send data this means that channel utilization is low and time slots assigned to neighboring nodes are wasted.

Z-MAC (Zebra MAC) protocol [3][17] is a kind of Frame-Slotted MAC protocols. The main drawback of TDMA is the low channel utilization as nodes have few data to send because only limited time is given to transmit data. Z-MAC is used to improve channel utilization by incorporating the CSMA into TDMA. Z-MAC applies DRAND (Distributed randomize) to do time slot assignment. DRAND ensures that no two nodes within the two-hop communication neighborhood are assigned to the same slot. There are various techniques to improve the channel utilization. TRAMA (Tree-search Auction Multiple Access) protocol is used by channel utilization by adaptive assignment. It switches between random access period and scheduled access period. In random access period nodes having data to send can only demand for slot while others cannot claim for the slot. NCR (Neighborhood-aware contention resolution) algorithm used to determine which node has slot is given by the formula in [3].

Priority(u, t)=hash(u+ t).

Where priority of u node at time t is found to be hash of concatenation of u and t. Node with high priority can win the slot. As it is based on TDMA time slot assignment incurs no communication overhead, the spatial reuse of time slots is low because nodes' priorities may be sequential. AI-LMAC [3] is also a protocol which adaptively assigns time slots to nodes when the sink initiates a query for data. Like D-MAC collision at sink node is a big issue. So to maximize the throughput at sink TreeMAC [3] protocol is used. It uses a time slot assignment algorithm that is well tuned for throughput maximization at sink by utilizing data gathering tree structure like in Fig. 4. The basic idea of TreeMAC is to eliminate horizontal two-hop interference by frame assignment and vertical two-hop interference by slot assignment. The transmission slot is calculated as $(L − 1) \mod 3$ where L is the depth of the node. Using three slots, a node can avoid contention with its previous and next hop. Still we are not achieving the low duty cycle with the above protocols. So to reduce the duty cycle Crankshaft [3] protocol is used which switches sending slots to receiving slots. In Earlier study TDMA assigns time slots to wake up intended node which saves energy to wake up other nodes. But there is no guarantee for collision free data transmission. In crankshaft time is divided into frames and frame further divided into slots. Each node listens for one unicast slot in every frame. A key feature of Crankshaft is that several broadcast slots are added to the end of unicast slots. Nodes can contend for broadcast in the broadcast slots. Crankshaft assigns unicast slots to nodes based on node ID modulo frame size. The simple assignment saves effort for learning schedules of neighbors, but two nodes may have the same unicast slot. Like crankshaft protocol PMAC [3] protocol also explained how time slot is assigned. Here care has been taken that sleeping time of node increased exponentially.

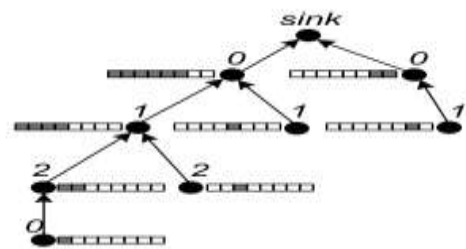

Fig. 4 Frame and Slot assignment in TreeMAC [3]

TABLE III
DIFFERENT FRAME-SLOTTED MAC PROTOCOLS [2] [3]

| Protocols | Technique | Advantage | Disadvantage |
|---|---|---|---|
| Z-MAC | Slot Stealing | Possesses high throughput under low contention. | Introduces additional overhead to detect abandoned slot. |
| TRAMA | Adaptive assignment | Channel utilization is more than Z-MAC | Spatial reuse of time slot is low because of sequential node priority. |
| TreeMAC | Maximize throughput at sink | Fairness is ensured in terms of flow instead of individual node. | A node needs a time to join the tree as it adopts CSMA also has lower channel priority. |
| Crankshaft | Reduce duty cycle by switching sending slots to receiving slots. | Only nodes have data can only wake up. | Time slots assigned to receiver hence collision may occur. |

## V. MULTICHANNEL MAC PROTOCOLS

Multichannel MAC protocols [3][10][12] is an area of interest for the wireless MAC protocol researchers because recent sensor platforms supports multiple channels. Fig. 5 can help to understand the need of multichannel design as here in this a single channel is not capable of doing parallel transmission. This will creates collision and interference among the channels [16]. In order to make fast information retrieval we need to imply parallel transmission mechanisms. So nodes will be assigned to different channels to increase the processing speed. Fig. 6 shows that how multiple channels are useful in order to send data simultaneously. By this technique we can send data high rate and achieve high throughput.

Similarly Fig. 7 explains the benefits of multiple channels over a single channel in traditional protocols. Single channel design needs 3 concurrent slots to send the data. While in multi-channel design parallel communications are possible [10][16]. The capacity of wireless networks increased by using multiple frequency bands. IEEE 802.11 standard play a vital role for contention based networks and divide the wireless spectrum into spectral bands called Channels [10][11][12]. Also one major reason for choosing multichannel protocols is the limited radio bandwidth in wireless sensor networks. With this limited bandwidth they need to support multitasking as well as to handle bursty traffic. Every researcher in WSN need to focus on efficient channel utilization and they have integrated various networks into a global advanced network.

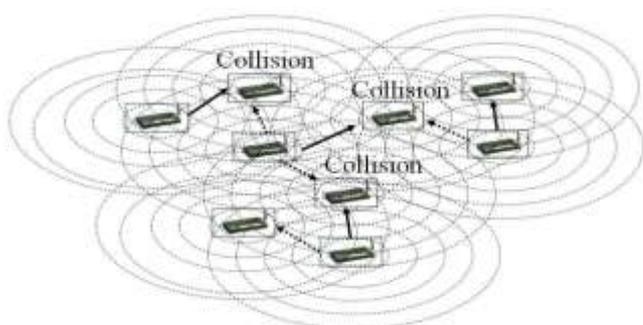

Fig. 5 Using a single channel increases collision [16]

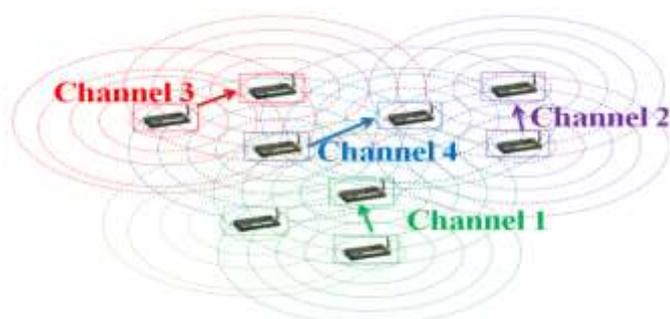

Fig. 6 Communication over Multiple channels [16]

First issue in this domain is the cross channel communication. Cross channel communication is always a challenging issue and can be solved by combining TDMA with FDMA to minimize the deficiency in cross channel. MMSN (Multi-Frequency MAC for WSN) [3][7][10][12] protocol is suitable for general wireless networks as well as ad hoc networks. They designed with the powerful radio and control overhead in more. MMAC [3] protocol assumes synchronized time which is further divided into fixed length beacon intervals. These beacons further divided into ATIM (Ad hoc traffic indication messages) and communication window. Nodes chooses the default channels in ATIM window for data transmission and then switch to communication window.

TMMAC [3] has same working like MMAC. Only they have made it flexible in the sense of data packet size. Hence they have used dynamic ATIM concept.

Channel assignment based on metric optimization [3] this technique suggest that frequency synthesizer needs time to stabilize hence more overhead observed in inter channel communication compared to the same channel. So they have suggested a method in which they have made a group of nodes which is frequently communicating into same channel while some nodes are not used regularly for the inter channel communication. They used k-way cut method to minimize the inter channel communication [10][12][16]. Simple idea behind this technique is to move some nodes to different channel when that channel is overloaded. TMCP[3][12][16] protocol partitions a sensor network to K vertex-disjoint trees, each of which is assigned a channel. Unlike K -way cut problem which minimizes inter-channel communication, TMCP tries to minimize the maximum intra-tree interference value among all trees. GBCA [3] protocol suggested that finding a channel assignment that minimizes total interference in a network is NP-hard problem.

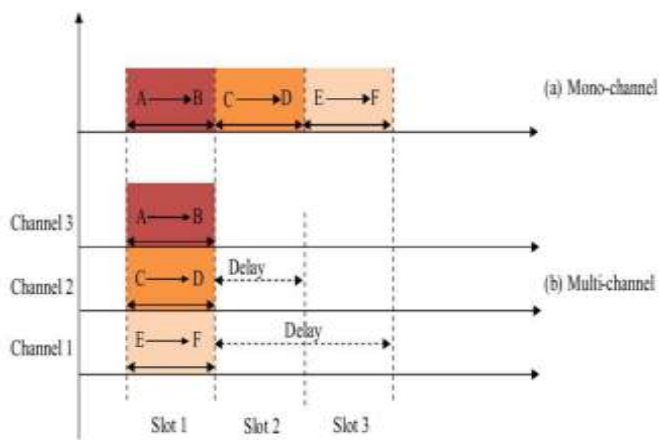

Fig. 7 Comparison between Single and Multi-channel Scheme [16]

If suppose frequent switching is carried out then it has to face additional overhead in terms of toggle transmission and toggle snooping. In MC-LMAC [3][10][12][16] protocol which uses TDMA/FDMA for sending technique time slot structure is used. MC-LMAC is a scheduled based protocol and follows semi-dynamic channel assignment method [15]. It categorizes the new coming node into a particular slot having common frequency. Same slot cannot occupied by a node within its two-hop communication when its neighbour has already occupied it. In TDMA/FDMA for receiving technique they noted that if a node has to send different amount of data in different periods. Sometimes it requires frequency hopping technique. Also it has to adopt dynamic traffic. The Y-MAC [3][14] protocol uses a dynamic channel selection method where time slot are assigned at receiver side. The time frame in Y-MAC consists of both broadcast as well as unicast period. The MuChMAC [3] is a hybrid design of TDMA and FDMA. Here receiving channel for each is autonomously chosen by each node. Hopping sequence is generated by pseudo random generator using node ID and slot number as an input.

TABLE IV
MULTICHANNEL MAC PROTOCOLS [3] [7][10][12][15][16]

| Protocol | Technique | Advantage | Disadvantage |
|---|---|---|---|
| MMSN | Cross channel communication | Parallel transmission is possible among neighboring nodes, | Powerful Radio size is not useful for WSN |
| TMMAC | Cross channel communication | Dynamic ATIM window to increase flexibility. | Node may loss energy by toggle snooping and toggle transmission |
| TMCP | Metric optimization | Minimize intra-tree interference value among all trees. | Metric does not reflect the actual interference intensity. |
| MC-LMAC | TDMA/FDMA for sending | No Frequent Channel switching. (Semi dynamic channel switching) | Overhead is slightly high and channel/slot utilization is low. |
| Y-MAC | TDMA/FDMA for receiving | Dynamic channel selection scheme introduced. | Contentions may occur when two receivers hop the same channel. |

## VI. OBSERVATION

We have discussed four categories of MAC protocols for wireless sensor network. Almost all categories have some design restrictions as they are application oriented [19]. Hence such methods cannot fit for all applications. Still we can say that multichannel MAC protocols can fulfil most of the requirements as we are in need of parallel processing or simultaneous transmission of data packets to improve the throughput. To design an energy efficient multichannel MAC protocol it has to save energy during channel switching. Also the channel selection mechanism in these protocols can allocate orthogonal as well non orthogonal i.e. overlapped channels [8] [9] [10].

## VII. CONCLUSION

In this paper we have discussed various categories of MAC protocols in wireless sensor networks. Mostly we have referred the classification from [3]. The classification carried out on the basis of factors like energy efficiency, delay, throughput etc.

To save the energy the node must be put into low power sleep mode is the basis for WSN MAC protocols. There are several techniques reviewed in this paper. To establish a connection Asynchronous MAC protocols introduced the prediction mechanism to estimate the best wake up time for sending. After reviewing some latest asynchronous techniques we can say that to establish a communication now responsibility goes to the receiver side. Synchronous MAC protocols techniques are vulnerable to interference. While in Frame-Slotted MAC protocols need to address contentions among multiple senders efficiently. Multichannel MAC protocols is the hot topic and having more challenges such as to design a dynamic channel allocation algorithm with low overhead. Efficient cross channel communication is one of the challenging issue. Multichannel designs need to improve the throughput by utilizing maximum orthogonal channels but we cannot guarantee the total utilization of channels. So non orthogonal or overlapped multichannel scheme may improve the throughput which is more than orthogonal channels throughput. From these all techniques we can say that all the methods with their protocols are application oriented designs. As there is no standard technique to classify all protocols with the same metrics.